\documentclass[10pt]{article} 

\begin{document}

\newcommand{\vk}{{\vec k}}
\newcommand{\vK}{{\vec K}} 
\newcommand{\vb}{{\vec b}} 
\newcommand{{\vp}}{{\vec p}} 
\newcommand{{\vq}}{{\vec q}} 
\newcommand{\vQ}{{\vec Q}}
\newcommand{\vx}{{\vec x}}
\newcommand{\beq}{\begin{equation}}
\newcommand{\eeq}{\end{equation}} 
\newcommand{\half}{{\textstyle \frac{1}{2}}} 
\newcommand{\gton}{\stackrel{>}{\sim}}
\newcommand{\lton}{\mathrel{\lower.9ex \hbox{$\stackrel{\displaystyle
<}{\sim}$}}} \newcommand{\ee}{\end{equation}}
\newcommand{\ben}{\begin{enumerate}} 
\newcommand{\een}{\end{enumerate}}
\newcommand{\bit}{\begin{itemize}} 
\newcommand{\eit}{\end{itemize}}
\newcommand{\bc}{\begin{center}} 
\newcommand{\ec}{\end{center}}
\newcommand{\bea}{\begin{eqnarray}} 
\newcommand{\eea}{\end{eqnarray}}
\newcommand{\beqar}{\begin{eqnarray}} 
\newcommand{\eeqar}[1]{\label{#1} \end{eqnarray}} \newcommand{\pleft}{\stackrel{\leftarrow}{\partial}}
\newcommand{\pright}{\stackrel{\rightarrow}{\partial}}

\begin{center}
{\Large {\bf{ Heavy Quark Radiative Energy Loss - Applications to RHIC}}}

\vspace{0.6cm}

{ Magdalena Djordjevic and Miklos Gyulassy }

\vspace{.3cm}

{\em { Dept. Physics, Columbia University, 538 W 120-th Street,New York,
       NY 10027, USA }} 

\vspace{.2cm}

\today

\end{center}

\vspace{.2cm}

\begin{abstract}
Heavy quark energy loss in a hot QCD plasma is computed taking into account the 
competing effects due to suppression of zeroth order gluon radiation bellow the 
plasma frequency and the enhancement of gluon radiation due to transition 
energy loss and medium induced Bremsstrahlung. Heavy quark medium induced 
radiative energy loss is derived  to all orders in opacity, $(L/\lambda_g)^n$. 
Numerical evaluation of the energy loss suggest small suppression of high 
$p_\perp$ charm quarks, and therefore provide a possible explanation for 
the null effects observed by PHENIX in the prompt electron spectrum in $Au+Au$ 
as $\sqrt{s}=130$ and $200$ AGeV.

\end{abstract}

\section{Introduction}

One of the most important goals of high energy heavy ion physics is the 
formation and observation of quark-gluon plasma (QGP). The discovery of a 
factor of $4\sim 5$ suppression of high $p_\perp \sim 5-10$ GeV hadrons 
produced in central $Au+Au$ at the Relativistic Heavy Ion Collider (RHIC) has 
been interpreted as evidence for jet quenching of light quark and gluon jets. 
Jet quenching was predicted~\cite{Gyulassy:1990bh} to occur due to radiative 
energy loss of high energy partons that propagate through ultra-dense QCD 
matter. The observed quenching pattern therefore provides a novel tomographic 
tool that can be used to map the evolution of the quark gluon plasma plasma 
(QGP) produced in ultra-relativistic nuclear collisions.

Though the observed jet quenching for light partons strongly suggest that 
QGP has been formed at RHIC, further detailed tests may provide the decisive 
proof for the discovery of QGP. It is believed that good test of 
QGP formation is the open charm suppression, which can now be  measured at 
RHIC by comparing $p_{\perp}$ distributions of $D$ mesons in $d+Au$ and 
$Au+Au$ collisions.

The main effect which lead to open charm suppression is the radiative heavy 
quark energy loss in a dense QCD medium. The first estimates for heavy quark 
energy loss~\cite{Shuryak:1996gc,Lin:1998bd} proposed that similar 
quenching may occur for charm jets as for light partons. However, in 
ref.~\cite{Dokshitzer:2001zm} it was estimated that the heavy quark mass 
leads to a kinematical ``dead cone'' effect for $\theta<M/E$ that reduces
significantly the induced radiative energy loss of heavy quarks. 
Experimentally, the PHENIX data~\cite{Adcox:2002cg} on ``prompt'' 
single electron production in $Au+Au$ collisions at $\sqrt{s}=130$ and 
$\sqrt{s}=200$ AGeV provided a first rough look at 
heavy quark transverse momentum distributions at RHIC.  Remarkably, no 
indication for a QCD medium effect was found within the admittedly large 
experimental errors. On the other hand, new STAR $D$ meson data may indicate 
large energy loss. Fortunately, in the near future, data with much higher 
statistics and wider $p_\perp$ range will become accessible.

In this proceedings we concentrate on the theory of heavy quark radiative 
energy loss, and how to use these results to present theoretical predictions 
that can be compared with upcoming experimental results. Such comparisons may 
provide decisive test for QGP existence.

Only the main results for heavy quark energy loss are presented here. For 
more detailed version see~\cite{DG_PLB,DG_PRC,DG_NPA}, and references therein.

\section{Heavy quark energy loss}

There are three important medium effects that control the radiative heavy 
quark energy loss in QCD matter. 

\medskip

Here we first study the non-abelian analog of the
Ter-Mikayelian~\cite{TM1} effect. The first estimates of the influence 
of a plasma frequency cutoff in QCD plasmas were reported in
ref.~\cite{Kampfer_Pavlenko} using a constant plasmon mass. In our study, we
extend those results by taking both longitudinal as well as transverse
modes consistently into account via the frequency and wavenumber
dependent hard thermal self energy~\cite{Gyulassy_Selikhov}. 

The detailed derivation of the QCD Ter-Mikayelian effect was presented in 
paper~\cite{DG_PRC}. An important conclusion from~\cite{DG_PRC} is that the 
effects due to the plasmon dispersion relation can be well approximated for 
high $p_\perp$ jets ignoring the longitudinal modes, and applying the 
asymptotic (short wavelength transverse) plasmon mass $m_g=\mu/\sqrt{2}$, 
where $\mu \approx g T$ GeV is chromoelectric Debye screening.

In addition to the Ter-Mikayelian effect,  we also need to take into account 
that medium has finite size. Additional radiation which occurs in 
the boundary between medium and vacuum has to be included. We will call this 
energy loss transition radiation. To estimate transition radiation we use the 
results from Zakharov~\cite{Zakharov}, which assume of static medium. This 
calculation should be improved by considering the more realistic case of 
expanding medium. 

On Fig.~1 we show the numerical results for a medium characterized by a Debye 
screening scale $\mu=0.5$ GeV. We see that the transverse plasmon mass effect
reduces the zeroth order energy loss by, $ \sim 30 \% $, relative to
the vacuum case. The gray region represents the additional energy loss which 
comes from transition radiation computed using~\cite{Zakharov}. This 
transition radiation lowers Ter-Mikayelian effect from $ \sim 30 \% $ to 
$ \sim 15 \% $. Therefore, these two effects would effectively {\em
enhance} the yield of high transverse momentum charm quarks were it
not for the extra medium induced radiation.

\begin{center}
\vspace*{4.5cm}
\includegraphics{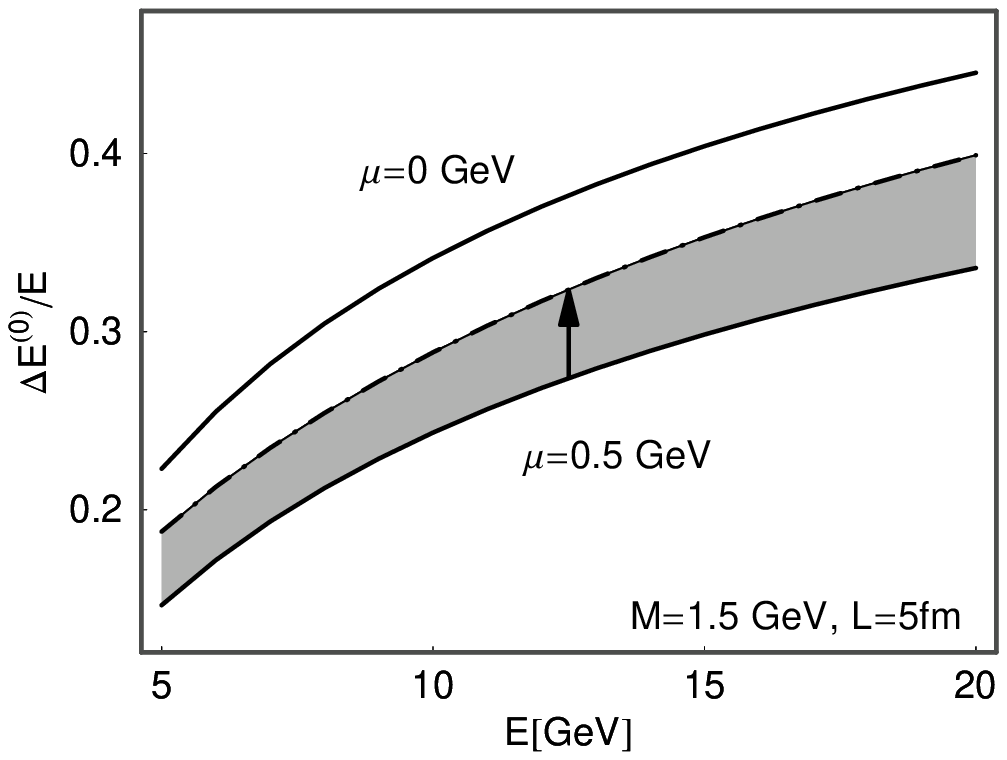}  
\begin{minipage}[t]{17.0cm}
{\small {FIG~1.} The reduction of the zeroth order (vacuum) energy loss for 
charm quark due to the QCD Ter-Mikayelian and transition radiation effects is 
shown as a function of the charm quark energy. The upper curve shows the 
vacuum energy loss if gluons are treated as massless and transversely 
polarized. The lower solid curve shows medium modified (but zeroth order in 
opacity) transverse fractional energy loss. The dot-dashed curve shows the 
additional effect of transverse radiation.}
\end{minipage}
\end{center}

The second part of our study is (1) to generalize the GLV opacity 
series~\cite{GLV} to include massive quark kinematic effects and (2) to take 
into account the Ter-Mikayelian plasmon effects for gluons as described 
in~\cite{DG_PRC}. The detailed study of this effect is 
presented in~\cite{DG_NPA}. We have derived heavy quark medium induced 
radiative energy loss to all orders in opacity, $(L/\lambda_g)^n$. The 
analytic expression generalizes the GLV opacity expansion for massless quanta
to heavy quarks with mass $M$ in a QCD plasma with a gluon dispersion
characterized by an asymptotic plasmon mass, $m_g=\mu/\sqrt{2}$. Remarkably, 
we find that the general result is obtained by simply shifting all 
frequencies in the GLV series by $(m_g^2+x^2 M^2)/(2 x E)$.

\begin{center}
\vspace*{4.5cm} 
\includegraphics{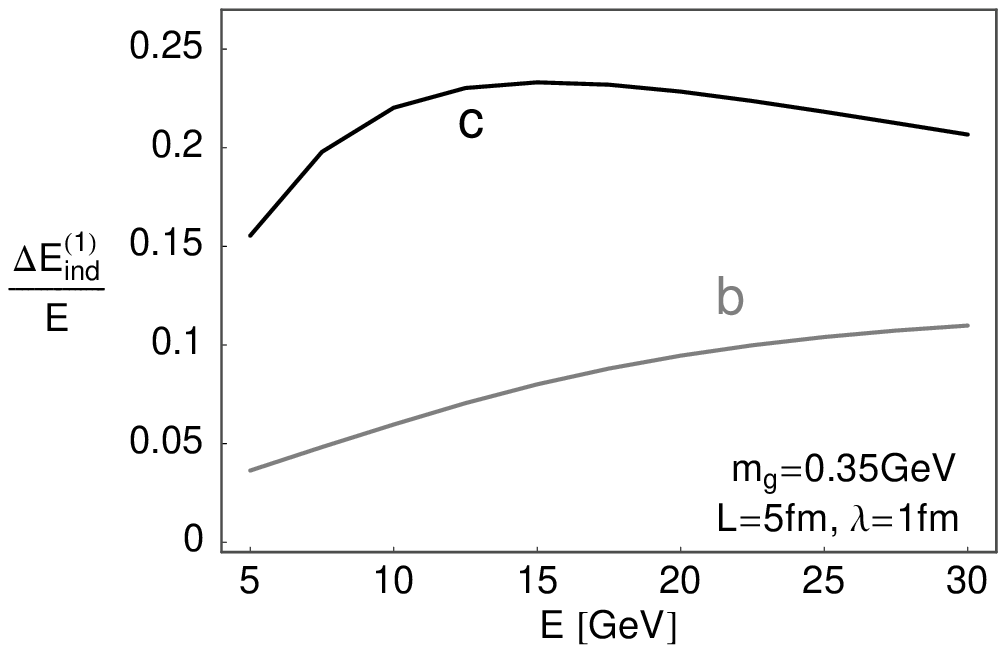}
\begin{minipage}[t]{17.cm}
{\small {FIG~2.} The $1^{st}$ order in opacity fractional energy loss for 
heavy quarks is shown as a function of their energy. Upper curve correspond 
to charm, and lower to bottom quarks in a plasma characterized by 
$\alpha_s=0.3, \; \mu=0.5\;{\rm GeV},$ and $L=5\lambda=5$ fm. 
}
\end{minipage}
\end{center}
\vskip 4truemm 

The numerical results for the first order induced radiative energy
loss are shown on Fig.~2 for charm and bottom quarks. We fix the
effective static plasma opacity to be $L/\lambda=5$. We see that, in the 
energy range $E \sim 5-15$ GeV, the 
induced energy loss fraction is $\Delta E^{(1)}/E \approx 0.2$ for charm 
quarks while only about half that is predicted for bottom.

\section{The net heavy quark energy loss}

Figure 3 shows the competition between the medium dependence of the induced 
energy loss and the zeroth order energy loss taking into account the 
Ter-Mikayelian effect and transition radiation. Even in the absence of a medium
($L=0)$, a charm quark with energy $\sim 10$ GeV suffers an average energy 
loss, $\Delta E^{(0)}_{vac}/E\approx 1/3$, due to the sudden change of the 
color current when it is formed in the vacuum. The dielectric plasmon effect 
reduces this to about $\Delta E^{(0)}_{med}/E\approx 1/4$. The additional 
transitional radiation lowers the difference between $\Delta E^{(0)}_{vac}$ 
and $\Delta E^{(0)}_{med}$.

\begin{center}
\vspace*{4.9cm} 
\includegraphics{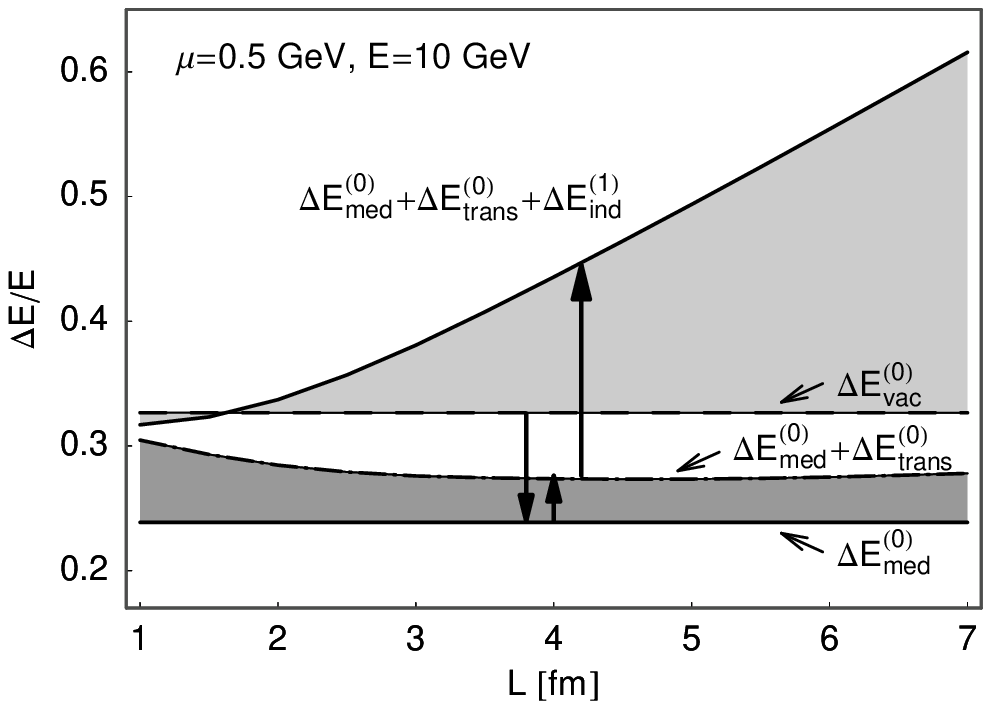}
\begin{minipage}[t]{17.cm}
{\small {FIG~3.} The fractional energy loss for a 10 GeV charm quark is 
plotted versus the effective static thickness $L$ of a plasma characterized 
by $\mu=0.5$ GeV and $\lambda=1$ fm. The dashed middle horizontal line 
corresponds to the energy loss in the vacuum taking into account the 
kinematic dead cone of radiation for heavy 
quarks~{\protect\cite{Dokshitzer:2001zm}}. The lower horizontal solid
line shows our estimate of the reduction of the zeroth order 
energy loss due to the QCD analog of the Ter-Mikayelian
effect. The dot-dashed middle curve shows the additional effect of transverse 
radiation. The upper solid 
curve corresponds to the net energy loss, $\Delta E^{(1)}_{ind}+
\Delta E^{(0)}_{med}+\Delta E^{(0)}_{trans}$. 
}
\end{minipage}
\end{center}

By comparing the net 
energy loss in the medium ($\Delta E^{(0)}_{med}+\Delta E^{(0)}_{trans}+
\Delta E^{(1)}_{ind}$) with the naive vacuum value ($\Delta E^{(0)}_{vac}$), 
we find that additional energy loss in the medium is approximately $15 \%$ 
of initial energy of the quark.

To estimate the value of suppression that comes from 
this energy loss we use the method described in~\cite{GLV_suppression}. We 
assume that initial charm $p_{\perp}$ distribution 
is in the region $\frac{C_{1}}{p_{\perp}^{8}}<\frac{d N^{in}}{d p_{\perp}}<
\frac{C_{2}}{p_{\perp}^{4}}$, where $C_{1}$ and $C_{2}$ are constants, and 
that the medium opacity is in the region $4.5-5.5$ fm.

\begin{center}
\vspace*{4.8cm} 
\includegraphics{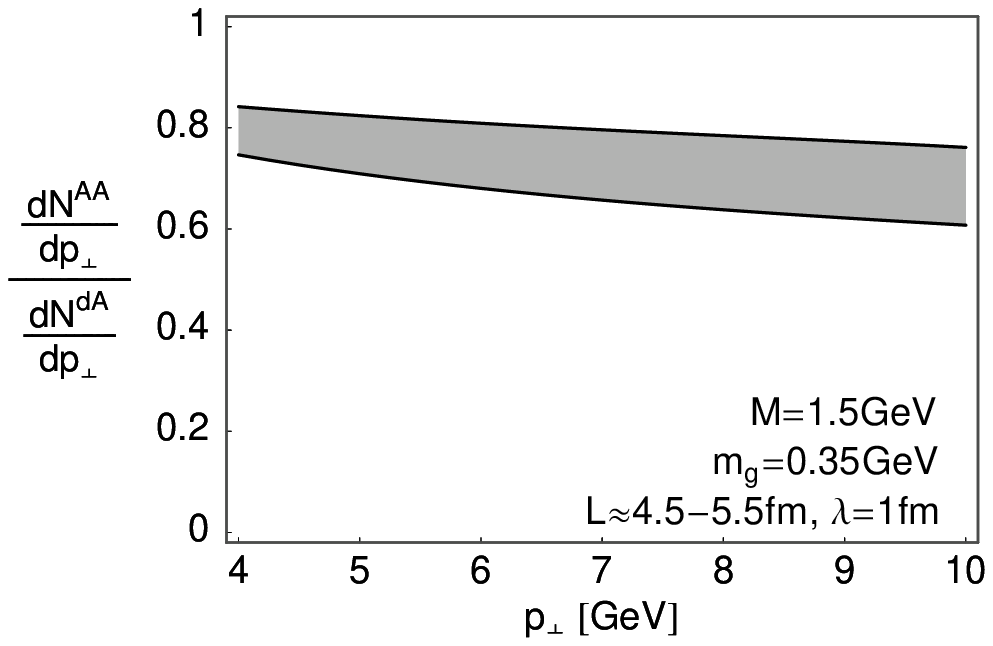}

\begin{minipage}[t]{9.cm}
{\small {FIG~4.} Charm $p_{\perp}$ suppression is shown as a function of 
$p_{\perp}$. 
}
\end{minipage}
\end{center}

From Fig.~4 we see that the charm quark suppression will be small 
$(0.6-0.8)$. Such value is expected having in mind low value 
of additional energy loss $(\approx 15\%)$. This prediction of suppression 
seems to be consistent with current PHENIX experimental single electron 
data~\cite{Adcox:2002cg}, but 
it is possibly inconsistent with STAR data. Unfortunately, the problem with 
current data is that they have large error bars, and that single electrons 
are less sensitive to the heavy quark energy loss than $D$ mesons. Therefore, 
high statistics $D$ meson data will allow us to make more reliable conclusions.

\section{Conclusions}

The upcoming $D$ meson data for 200 GeV $d+Au$ and $Au+Au$ results will 
soon become available. According to our results, charm quark suppression 
should be small $( 0.6-0.8)$. Therefore, this suppression should be definitely 
much smaller than the already observed pion suppression $(0.2)$.
 
If our predictions are confirmed, then existence of QGP will pass 
another stringent test. Together with already observed jet quenching and 
elliptic flow of light partons, this may provide decisive argument in the 
favor of the QGP production at RHIC.

\vspace{0.5 cm}
{\em Acknowledgments:} We would like to thank R. Vogt for carefully reading 
the paper and giving valuable suggestions. This work is supported by the 
Director, Office of Science, Office of High Energy and Nuclear Physics, 
Division of Nuclear Physics, of the U.S. Department of Energy under Grant No.
DE-FG02-93ER40764.

\end{document}